\documentstyle[psfig]{article}

\begin{document}

\noindent {\it I.~F.~Schegolev's Memorial Volume,
Journal de Physique I (France) {\bf 6} (1996) 1917} 
\bigskip\bigskip

\noindent PACS: 73.40.Hm --- 72.15.Nj --- 72.80.Le

\bigskip\bigskip

\noindent
{\Large\bf Quantum Hall Effect in Quasi-One-Dimensional Conductors:
The Roles of Moving FISDW, Finite Temperature, and Edge States }
\bigskip

\hfill
\begin{minipage}{13.5cm}

\noindent
Victor M.\ Yakovenko\footnotemark\
and Hsi-Sheng Goan\footnotemark
\bigskip

\noindent
Department of Physics and Center for Superconductivity Research,
University of Maryland, College Park, MD 20742-4111, USA
\bigskip

\noindent
{\bf cond-mat/9607129, first submitted 17 July 1996, revised January 21, 1997}
\bigskip

\noindent

   {\bf Abstract.} --- This paper reviews recent developments in the
theory of the quantum Hall effect (QHE) in the magnetic-field-induced
spin-density-wave (FISDW) state of the quasi-one-dimensional organic
conductors (TMTSF)$_2$X.  The origin and the basic features of the
FISDW are reviewed.  The QHE in the pinned FISDW state is derived in
several simple, transparent ways, including the edge states
formulation of the problem.  The temperature dependence of the Hall
conductivity is found to be the same as the temperature dependence of
the Fr\"ohlich current.  It is shown that, when the FISDW is free to
move, it produces an additional contribution to the Hall conductivity
that nullifies the total Hall effect.  The paper is written on
mathematically simple level, emphasizes physical meaning over
sophisticated mathematical technique, and uses inductive, rather than
deductive, reasoning.

\bigskip

\end{minipage}
\addtocounter{footnote}{-1}
\footnotetext{E-mail: yakovenk@glue.umd.edu}
\stepcounter{footnote}
\footnotetext{E-mail: goan@glue.umd.edu}

\section{Introduction}
\label{Sec:Intro}

   Organic metals of the (TMTSF)$_2$X family, where TMTSF is
tetramethyltetraselenafulvalene and X represents an inorganic anion,
are highly anisotropic, quasi-one-dimensional (Q1D) crystals that
consist of parallel conducting chains.  The overlap of the electron
wave functions and the electric conductivity are the highest in the
direction of the chains (the {\bf a} direction) and are much smaller
in the {\bf b} direction perpendicular to the chains.  In this paper,
we neglect the coupling between the chains in the third, {\bf c}
direction, which is weaker than in the {\bf b} direction.  We study
the properties of a single layer (the {\bf a}-{\bf b} plane) that
consist of weakly coupled parallel chains, modeling (TMTSF)$_2$X as a
system of the uncoupled two-dimensional (2D) layers.

   The (TMTSF)$_2$X materials exhibit very interesting behavior when a
strong magnetic field is applied perpendicular to the {\bf a}-{\bf b}
plane.  At low temperature below several Kelvin and magnetic field of
the order of five Tesla, there is a phase transition from the metallic
state to a state, where the spin-density wave appears.  This state is
called the magnetic-field-induced spin-density-wave (FISDW) state (see
Ref.\ \cite{Yamaji90} for a review).  As the magnetic field is
increased further, a sequence (cascade) of phase transitions between
different FISDWs is observed.  Interestingly, within each FISDW phase,
the value of the Hall resistance remains constant, independent of the
magnetic field, that is, the quantum Hall effect (QHE) is observed.
Once the boundary to another FISDW phase is crossed, the value of the
Hall resistance jumps to a new value, which remains constant until the
next phase boundary is crossed.

   It is instructive to compare the QHE in the FISDW state with the
conventional QHE observed in 2D semiconductor devices.  In both cases,
at sufficiently low temperatures, the longitudinal resistivity
$\rho_{xx}$ is much smaller than the Hall resistivity $\rho_{xy}$ and is
thermally activated with an energy gap, which is equal to 6 K in one of
the FISDW phases \cite{Chaikin88d}.  The theory says that the Hall
conductivity per one layer is quantized: $\sigma_{xy}=2Ne^2/h$.  Whereas
in the case of a 2D electron gas in a single semiconducting layer this
formula can be directly verified experimentally, in the (TMTSF)$_2$X
materials the situation is more complicated.  What is measured
experimentally in (TMTSF)$_2$X is the total, bulk Hall resistance of
many parallel layers.  To find the Hall conductivity per one layer, one
needs to know the effective number of layers contributing to the Hall
conductivity, which depends on the electric current distribution in the
sample and, thus, is somewhat uncertain.  So, in (TMTSF)$_2$X, one can
only compare the relative values of the Hall resistances at different
plateaus and deduce the integer numbers $N$ from these ratios.  For this
reason, it is hard to unambiguously discriminate experimentally between
the integer and the fractional QHE in the (TMTSF)$_2$X materials.
Nevertheless, the common belief, strongly influenced by the theory (see
the rest of the paper), is that the QHE in (TMTSF)$_2$X is the integer
one.

   Unlike semiconductors, the (TMTSF)$_2$X materials have very high,
metallic concentration of carriers (one conducting hole per unit
cell).  Thus, in (TMTSF)$_2$X, a naively calculated filling factor of
the Landau levels in a realistic magnetic field is enormous, of the
order of $10^2-10^3$, depending on the magnitude of the field.  At the
same time, the Hall conductivity is quantized with a small,
single-digit number $N$.  The discrepancy between the naive filling
factor and the value of the Hall conductivity is resolved by the very
important fact that the QHE in (TMTSF)$_2$X exists solely due to the
phase transition into a FISDW state.  The FISDW effectively eliminates
most of the carriers, reducing the filling factor to the single-digit
number $N$, which manifests itself in the value of the Hall
conductivity.  In this respect, the QHE in the (TMTSF)$_2$X materials
significantly differs from the conventional QHE in semiconductors,
where the QHE state is not associated with any thermodynamic phase
transition and order parameter.  In (TMTSF)$_2$X, the transitions
between the QHE plateaus are true thermodynamic phase transitions,
accompanied by changes in the FISDW order parameter and observed in
the measurements of specific heat \cite{Pesty85}, magnetization
\cite{Chaikin85}, NMR \cite{Takahashi84}, and virtually any other
physical quantity.  For a given magnetic field, the effective filling
factor $N$ is determined by delicate and nontrivial FISDW
thermodynamics, which may vary from one material to another.  Thus,
the ratios of the Hall resistances in the consecutive FISDW phases of
(TMTSF)$_2$PF$_6$ are typically equal to the ratios of the consecutive
integer numbers 1:2:3:4:5 \cite{Jerome89}, whereas in
(TMTSF)$_2$ClO$_4$ \cite{Chaikin88a} and (TMTSF)$_2$ReO$_4$
\cite{Jerome91b} the ratios do not follow any simple sequence and may
change sign.  In this paper, we do not discuss the FISDW
thermodynamics in detail.  In Sec.\ \ref{Sec:FISDW} we only
demonstrate that a FISDW is characterized by an integer number $N$,
and in Sec.\ \ref{Sec:QHE} we show that this number appears in the
expression for the Hall conductivity.  However, we do not calculate
how the number $N$ depends on the magnetic field and other parameters
of the model.  These issues are discussed in detail in the theory of
the FISDW formation (see Refs.\ \cite{Yamaji90,Montambaux91} for
reviews).

   Early theoretical approaches \cite{Montambaux84b,Chaikin87a}
explained the QHE in the FISDW state by counting the number of
carriers left after the FISDW gap opens.  While this calculation gives
correct answer, it is not completely satisfactory, because the FISDW
gap is much smaller than the cyclotron frequency of the magnetic
field, which makes the FISDW case totally opposite to the standard
semiconductor situation, from where the concept of the calculation is
borrowed.  Furthermore, the ``insulating'' FISDW gap and the
``Landau'' gaps due to the magnetic field hybridize very strongly,
which make the situation even more complicated. In Ref.\
\cite{Montambaux87}, the QHE was derived rigorously, albeit somewhat
indirectly, using the Streda formula.  In Ref.\ \cite{Yakovenko91a},
the QHE was calculated directly, using a manifestly
topologically-invariant expression for the Hall conductivity in terms
of the electrons wave functions that follows from the Kubo formula.
This approach can be straightforwardly generalized
\cite{Yakovenko91a,Yakovenko94a} to the case where several FISDW order
parameters coexist \cite{Lebed90}.

   The present paper is devoted mostly to recent developments in the
theory of the QHE in the FISDW state.  In Sec.\ \ref{Sec:FISDW}, we
explain the basics of the FISDW.  In Sec.\ \ref{Sec:QHE}, we give yet
another derivation of the QHE that emphasizes analogy between the QHE
and the Fr\"ohlich conduction of a charge/spin-density wave.  In Sec.\
\ref{Sec:Moving}, this analogy is utilized to discuss what happens to
the QHE when the FISDW moves.  In Sec.\ \ref{Sec:Temperature}, the
effect of a finite temperature on the QHE is calculated.  In Sec.\
\ref{Sec:Edges}, we reformulate the QHE in terms of the edge states.
In Sec.\ \ref{Sec:Conclusion}, conclusions are given.  Throughout the
paper, we try to keep discussion on mathematically simple level,
emphasizing physical meaning over sophisticated mathematical technique
and using inductive, rather than deductive, reasoning.

\section{Formation of the FISDW }
\label{Sec:FISDW}

   For pedagogical purposes, let us start consideration from a simple
one-dimensional (1D) system, where electrons are confined to a chain
parallel to the $x$ axis.  Suppose the electron dispersion law is
parabolic, so the Hamiltonian $\hat{\cal H}$ can be written
as:\footnote{The actual form of the longitudinal dispersion law
(\ref{Hm}) is not very essential.}
\begin{equation}
\hat{\cal H}=\hbar^2 k_x^2/2m,
\label{Hm}
\end{equation}
where $\hbar=h/2\pi$ is the Planck constant, $m$ is the electron mass,
and $k_x$ is the electron wave vector along the chain.  At zero
temperature, the electrons occupy the quantum states with the wave
vectors from $-k_{\rm F}$ to $k_{\rm F}$ and the energies up to
$E_{\rm F}$, where $k_{\rm F}$ and $E_{\rm F}$ are the Fermi wave vector
and energy, which are determined by the concentration of the
electrons.

   Now, suppose that a periodic potential is present in the system, so
that the Hamiltonian is equal to:
\begin{equation} 
\hat{\cal H}=-\frac{\hbar^2}{2m}\frac{\partial^2}{\partial x^2} +
2\Delta\cos(Q_x x),
\label{HD} 
\end{equation} 
where $Q_x$ is the wave vector and $\Delta\ll E_{\rm F}$ is the
amplitude of the periodic potential.  As it is well known from quantum
mechanics, the periodic potential opens an energy gap of the magnitude
$2\Delta$ in the electron spectrum at the wave vectors $k_x=\pm
Q_x/2$.\footnote{Smaller gaps, opened at the higher integer multiples
of $\pm Q_x/2$, are not essential for our consideration.} If the wave
vector of the periodic potential connects the two Fermi points of the
electrons:
\begin{equation} Q_x=2k_{\rm F},
\label{Q0} 
\end{equation} 
the gap opens right at the Fermi level, so the states below the gap
are completely occupied and the states above are completely empty.  It
is clear that the total energy of the electrons is reduced compared to
the total energy in the absence of the periodic potential.  Thus, if
the electrons interact between themselves, they might decide to
produce the periodic potential spontaneously, self-consistently in
order to reduce the total energy of the system.  This phenomenon is
called the Peierls instability.  Once the periodic potential appears
in the system, it modulates the charge or spin density of the
electrons, producing a charge- or spin-density wave (CDW/SDW) with the
wave vector $Q_x$ (see Ref.\ \cite{Gruner94b} for a review).  We do
not discuss here the details of the interaction between the electrons
that leads to formation of the periodic potential.  In this paper, we
focus only on the mean-field periodic potential experienced by the
electrons once the CDW/SDW has been established, presuming that the
self-consistency conditions are satisfied.  For our purposes, the
distinction between the CDW and SDW is not important, so we pay no
attention to the spin indices.

   Now, to model (TMTSF)$_2$X, let us consider a 2D system that
consists of many chains, parallel to the $x$ axis and equally spaced
along the $y$-axis with the distance $b$.\footnote{The $x$ and $y$
axes correspond to the {\bf a} and {\bf b} axes of (TMTSF)$_2$X.} The
chains are coupled through the electron tunneling of the amplitude
$t_b$, so the electron Hamiltonian is:
\begin{equation}
\hat{\cal H}=-\frac{\hbar^2}{2m}\frac{\partial^2}{\partial x^2} +
2\Delta\cos(Q_x x) + 2t_b\cos(k_y b),
\label{Ht}
\end{equation}
where $k_y$ is the electron wave vector across the chains.
Hamiltonian (\ref{Ht}) is written in the mixed representation, where
an electron wave function depends on the coordinate $x$ along the
chains and the momentum $k_y$ across the chains.  As follows from Eq.\
(\ref{Ht}), the electron energy now depends on the momentum $k_y$.
Strictly speaking, in the presence of many chains, the CDW/SDW
potential may also have a certain periodicity across the chains and
should be written as $2\Delta\cos(Q_x x+Q_y nb)$, where $n$ is the
chain number and $Q_y$ is the wave vector of the CDW/SDW across the
chains.  To simplify calculations, we consider only the case of
$Q_y=0$, which is not the most realistic case, but the results are
qualitatively valid also in a more realistic case of $Q_y\neq 0$.  To
achieve quantitative agreement between the theory and experiment, it
may be necessary to consider a more complicated transverse dispersion
law of the electrons and to include the next-nearest-neighbor hopping
term $2t_b'\cos(2k_y b)$ in the Hamiltonian.  To simplify out
qualitative discussion, we neglect this term.

   Now, suppose that a magnetic field $H$ is applied along the
$z$-axis perpendicular to the $(x,y)$-plane. To describe the magnetic
field, we select the Landau gauge:\footnote{The fact that we use a
specific gauge does not invalidate our results in any way. This gauge
is selected to simplify calculations. We can perform the calculations
in the most general, arbitrary gauge, but the formulas would be more
complicated.}
\begin{equation}
A_x=A_z=0,\quad A_y=Hx,
\label{A}
\end{equation}
and do the Peierls--Onsager substitution, $k_y\rightarrow
k_y-eA_y/c\hbar$, in Hamiltonian (\ref{Ht}). The Hamiltonian becomes:
\begin{equation}
\hat{\cal H}=-\frac{\hbar^2}{2m}\frac{\partial^2}{\partial x^2} +
2\Delta\cos(Q_x x) + 2t_b\cos(k_y b-G_x x),
\label{HH}
\end{equation}
where
\begin{equation}
G_x=ebH/\hbar c.
\label{Gx}
\end{equation}
Comparing Eqs.\ (\ref{Ht}) and (\ref{HH}), we see that, in the
presence of the magnetic field, the hopping {\em across} the chains
becomes a periodic potential {\em along} the chains with the wave
vector $G_x$ (\ref{Gx}).  We will refer to this periodic potential as
the ``hopping potential''.  The period of this potential (the magnetic
length),
\begin{equation}
l_H=2\pi/G_x,
\label{lH}
\end{equation}
is determined by the condition that magnetic flux through a
2D cell formed by the magnetic length along the chains,
$l_H$, and the distance between the chains, $b$, is equal to the flux
quantum, $\phi_0$:
\begin{equation}
l_HbH=\phi_0=hc/e.
\label{phi0}
\end{equation}
Eq.\ (\ref{phi0}) is equivalent to Eqs.\ (\ref{Gx}) and (\ref{lH}).

   According to Eq.\ (\ref{HH}), in the presence of both the CDW/SDW
and the magnetic field, the electrons experience two periodic
potentials with the wave vectors $Q_x$ and $G_x$.  The magnitudes of
the two wave vectors are very different. $Q_x$ is big, of the order of
$2k_{\rm F}$, and the corresponding period, $l_{\rm DW}=2\pi/Q_x$, is
short, of the order of the distance between the electrons.  On the
other hand, in realistic magnetic fields, the magnetic length $l_H$ is
much longer than the inter-electron distance, thus the ratio of the
wave vectors, $G_x/Q_x$, is very small, of the order of
$10^{-2}-10^{-3}$, depending on the value of the magnetic field.
Thus, the two periodic potentials can be treated as incommensurate.
In this case, the energy spectrum is degenerate in $k_y$, because
changing $k_y$ means simply shifting the hopping potential in Eq.\
(\ref{HH}) along the $x$ axis, which does not change the energy.

   To get qualitative picture of the energy spectrum produced by the
two periodic potentials, let us assume for a moment that $t_b$ is very
small and can be treated as a perturbation. Taken alone, the CDW/SDW
potential opens a gap in the electron spectrum at the wave vectors
$k_x=\pm Q_x/2$ connected by the CDW/SDW wave vector $Q_x$.  The
CDW/SDW potential, combined perturbationally with the hopping
potential, opens gaps at the wave vectors $k_x=\pm(Q_x\pm G_x)/2$
connected by the wave vectors $Q_x\pm G_x$ obtained by combining the
wave vectors of the two periodic potentials. In the same manner, the
CDW/SDW potential, combined $n$ times with the hopping potential,
opens gaps at the wave vectors $k_x=\pm(Q_x\pm nG_x)/2$ connected by
the combinational wave vectors $Q_x\pm nG_x$. Thus, the electron
spectrum contains a sequence of energy gaps, which are equally spaced
in momentum $k_x$ with the distance $G_x/2$ \cite{Montambaux86}. The
gaps separate energy bands; each band has the total width $\Delta
k_x=G_x$. These bands can be interpreted as the Landau levels
broadened into the energy bands (with the dispersion in $k_x$) by the
periodic arrangement of the chains with the period $b$. The Landau
degeneracy in $k_y$ remains in the problem.\footnote{When several
FISDW order parameters with different wave vectors (\ref{Qx}) coexist,
the degeneracy in $k_y$ is lifted.}  The number of states per unit
area in each band is equal to
\begin{equation}
\frac{\Delta k_x\,\Delta k_y}{(2\pi)^2}=
\frac{G_x}{(2\pi)^2}\frac{2\pi}{b}=\frac{eH}{hc},
\label{DkxDky}
\end{equation}
which coincides with the number of states in a Landau level.

   In the (TMTSF)$_2$X materials, the interchain hopping is not that
small and, generally speaking, cannot be treated as a
perturbation.\footnote{See Sec.\ \ref{Sec:Temperature} for a
nonperturbative treatment of the problem.} Nevertheless, the
qualitative picture of the electron energy spectrum outlined above
remains valid with the important quantitative difference that some
``secondary'' gaps, opened at the combinational wave vectors
$Q_x+nG_x$, may be bigger than the ``primary'' gap, opened at $Q_x$.
Since the CDW/SDW potential is produced self-consistently to maximize
the energy gain, the electrons would create the CDW/SDW with such a
wave vector that the biggest secondary energy gap is located exactly
at the Fermi level. In this case, the wave vector of the biggest gap,
$Q_x+NG_x$, characterized by some integer number $N$, must coincide
with $2k_{\rm F}$, the span of the Fermi sea: $Q_x+NG_x=2k_{\rm
F}$. Thus, the wave vector of the CDW/SDW is determined by the
following equation:
\begin{equation}
Q_x=2k_{\rm F}-NG_x=2k_{\rm F}-NebH/\hbar c.
\label{Qx}
\end{equation}
This is the most important equation of this Section. It shows that, in
a multichain, 2D system subject to a magnetic field, the longitudinal
wave vector of the CDW/SDW is not necessarily equal to $2k_{\rm F}$,
as it was in strictly 1D system (\ref{Q0}), but may take many
different values (\ref{Qx}) labeled by an integer number $N$
\cite{Montambaux84b}. In this paper, we do not calculate which values
of $N$ the system selects for a given magnetic field $H$ and given
microscopic parameters of the model ($t_b$, $E_{\rm F}$, the amplitude
of interaction between the electrons $g$, etc.). These issues are
addressed in reviews \cite{Yamaji90,Montambaux91}, as well as in
original articles, e.g.\
\cite{Montambaux84b,Yakovenko91a,Lebed85,Montambaux85,Montambaux86}. We
assume that the value of $N$ is given to us and study the properties
of the system in this state, specifically the Hall effect.

   The CDW/SDW wave vector (\ref{Qx}) changes linearly with the magnetic
field $H$ in order to keep the energy gap exactly at the Fermi level.
Because the magnetic field is intrinsically involved in the formation of
the energy gap at the Fermi level, this kind of density wave in
(TMTSF)$_2$X is called the magnetic-field-induced spin-density wave
(FISDW).\footnote{The density wave happens to be the spin, not the
charge one in (TMTSF)$_2$X, which is not essential for our discussion.}
We will use the term FISDW in the rest of the paper.

\section{The Quantum Hall Effect}
\label{Sec:QHE}

   Let us discuss the Hall conductivity of our 2D system in the FISDW
state at zero temperature. By naive analogy with conventional
semiconductors, one might say \cite{Montambaux84b,Chaikin87a} that all
electron states below the ``primary'' gap, opened by the FISDW
potential at the wave vector $Q_x$, do not contribute to the Hall
conductivity. Thus, the effective number of carriers per one chain is
the difference between the total number of carriers, proportional to
the size of the Fermi sea $2k_{\rm F}$, and the number of the
``eliminated'' carriers, proportional to $Q_x$:
\begin{equation}
\rho_{\rm eff}=2\,\frac{2k_{\rm F}-Q_x}{2\pi b}=\frac{2NeH}{hc},
\label{rhoeff}
\end{equation}
where the first factor 2 comes from the spin, and the second equality
follows from Eq.\ (\ref{Qx}). Substituting Eq.\ (\ref{rhoeff}) into the
conventional formula for the Hall conductivity:
\begin{equation}
\sigma_{xy}=\rho_{\rm eff}ec/H,
\label{rec/H}
\end{equation}
we see that the magnetic field cancels out and the Hall
conductivity is quantized:
\begin{equation}
\sigma_{xy}=2Ne^2/h.
\label{2Ne2/h}
\end{equation}
This derivation can be summarized as follows. The FISDW wave vector
(\ref{Qx}) adjusts its value to the magnetic field in such a manner
that there are always $N$ completely filled Landau bands between the
``primary'', ``insulating'' FISDW gap and the Fermi level. Thus, the
Hall conductivity is quantized with the effective number of the Landau
bands $N$. It is by the elimination of almost all of the carriers the
FISDW reduces the effective filling factor from $10^2-10^3$ to the
single-digit number $N$.

   Although the above derivation of $\sigma_{xy}$ gives correct answer
(\ref{2Ne2/h}), it raises many questions and doubts. Why do we say
that the ``primary'' gap, which is even not the biggest one,
``eliminates'' the carriers from the Hall effect, whereas the
``secondary'' gaps do not? Is formula (\ref{rec/H}) applicable in our
situation? To our opinion, the derivation given above is not
convincing, and below we give another, rigorous derivation, which is
based on the ideas of Refs. \cite{Zak85,Kunz86}.

   Suppose the electric field $E_y$ is applied perpendicular to the
chains. Let us use the following gauge
\begin{equation}
A_x=A_z=0,\quad A_y=Hx-E_yct,
\label{At}
\end{equation}
where $t$ is the time. In the presence of the electric field, the
electron Hamiltonian (\ref{HH}) becomes
\begin{equation}
\hat{\cal H}=-\frac{\hbar^2}{2m}\frac{\partial^2}{\partial x^2} +
2\Delta\cos(Q_x x+\Theta) + 2t_b\cos[k_y b-G_x(x-v_{E_y}t)],
\label{Ham}
\end{equation}
where, for further purposes, we introduced an arbitrary phase $\Theta$
in the FISDW potential, and
\begin{equation}
v_{E_y}=cE_y/H
\label{vEy}
\end{equation}
is the drift velocity in the crossed electric and magnetic fields.

   We see that, in the presence of the transverse electric field
$E_y$, the hopping potential in Eq.\ (\ref{Ham}) moves along the
chains with the velocity $v_{E_y}$ (\ref{vEy}) proportional to $E_y$.
Because in the FISDW state all electrons are under the energy gap, by
analogy with the Fr\"ohlich conduction produced by motion of a
CDW/SDW, the motion of the hopping potential in Eq.\ (\ref{Ham})
should induce some electric current along the chains, $j_x$,
proportional to the velocity $v_{E_y}$. This is the Hall current, and,
once we know $j_x$, the Hall conductivity can be calculated as
$\sigma_{xy}=j_x/E_y$. The difficulty of our problem is that there are
two different periodic potentials in Hamiltonian (\ref{Ham}), due to
the FISDW and due to the hopping. Normally, the FISDW potential is
pinned and does not move, whereas the hopping potential moves due to
the presence of the electric field $E_y$ in its argument and cannot be
pinned.

   In order to calculate $\sigma_{xy}$, let us consider a more general
case where the FISDW potential may also move. To find the Hall
conductivity of the pinned FISDW, we will set the FISDW velocity to
zero at the end of the calculation. According to Eq.\ (\ref{Ham}),
when the FISDW potential moves, the FISDW phase $\Theta$ changes in
time $t$, so that the velocity of the motion $v_{\rm DW}$ is
proportional to the time derivative $\dot{\Theta}$:
\begin{equation}
v_{\rm DW}=-\dot{\Theta}/Q_x.
\label{vDW}
\end{equation}
We assume that both $E_y$ and $\dot{\Theta}$ are infinitesimal, thus
the velocities $v_{E_y}$ (\ref{vEy}) and $v_{\rm DW}$ (\ref{vDW}) are
very small, so the motion of the potentials is adiabatic.

   Now, let us calculate the current along the chains produced by the
motion of the potentials. Since there is an energy gap at the Fermi
level, following the arguments of Laughlin \cite{Laughlin81}, we can
say that an integer number of electrons $N_1$ is transferred from one
end of a chain to another, when the FISDW potential is adiabatically
shifted along the chain by its period $l_{\rm DW}=2\pi/Q_x$.  The same
is true, with an integer $N_2$ instead of $N_1$, for a displacement of
the hopping potential by its period $l_H=2\pi/G_x$.  Because the two
potentials are incommensurate, if the first potential is shifted by
$dx_1$ and the second by $dx_2$, the total transferred charge $dq$ is
the sum of the prorated amounts of $N_1$ and $N_2$:
\begin{equation}
dq=eN_1\frac{dx_1}{l_{\rm DW}}+eN_2\frac{dx_2}{l_H}.
\label{dq}
\end{equation}
Now, suppose that both potentials are shifted by the same displacement
$dx=dx_1=dx_2$. In this case, we can also write that
\begin{equation}
dq=e\rho\,dx,
\label{rho}
\end{equation}
where $\rho=4k_{\rm F}/2\pi$ is the concentration of the electrons. Equating
(\ref{dq}) and (\ref{rho}) and substituting the expressions for
$\rho$, $l_{\rm DW}$ (\ref{Qx}), and $l_H$ (\ref{lH}), we find the
following Diophantine-type equation \cite{Zak85}:
\begin{equation}
4k_{\rm F}=N_1 (2 k_{\rm F}-NG_x)+N_2 G_x,
\label{Diophant}
\end{equation}
where $N$ is the integer that characterizes the FISDW. Since $k_{\rm F}/G_x$
is, in general, an irrational number, the only possible solution of
Eq.\ (\ref{Diophant}) for the integers $N_1$ and $N_2$ is
\begin{equation}
N_1=2, \quad\quad N_2=N_1 N=2N.
\label{N12}
\end{equation}

   Dividing Eq.\ (\ref{dq}) by the distance between the chains $b$ and
by the time increment $dt$ and using expressions (\ref{vEy}) and
(\ref{vDW}) for the velocities and (\ref{N12}) for the integers, we
find the density of current along the chains:
\begin{equation}
j_x=-\frac{e}{\pi b}\dot{\Theta} + \frac{2Ne^2}{h}E_y.
\label{jx}
\end{equation}
The first term in Eq.\ (\ref{jx}) represents the contribution of the
FISDW motion, the so-called Fr\"{o}hlich conductivity
\cite{Gruner94b}. This term vanishes when the FISDW is pinned and does
not move ($\dot{\Theta}=0$). The second term describes the quantum
Hall effect. The expression for $\sigma_{xy}$ that follows from Eq.\
(\ref{jx}) coincides with Eq.\ (\ref{2Ne2/h}). Apparently, the QHE in
the FISDW is the integer one, and the derivation given above seems to
exclude a possibility of the fractional QHE.

   Having derived the QHE in the FISDW state, let us compare it with
the conventional integer QHE in semiconductors. In the latter systems,
the electron localization due to disorder is thought to play an
important role by providing a reservoir of electron states necessary
to maintain a constant value of the Hall conductivity with the varying
magnetic field. On the other hand, in the FISDW state, we deal with
the QHE in a clean, periodic 2D potential.  This problem was
considered in Ref.\ \cite{Thouless}, either for a tight-binding model,
or for two weak sinusoidal potentials in the $x$ and $y$
directions. In the FISDW state, we have an intermediate case, where
the chains produce a tight-binding potential in the $y$ direction,
whereas the FISDW provides a weak sinusoidal potential in the $x$
direction.\footnote{The crystal lattice periodicity in the $x$
direction does not play essential role in our model and may be
neglected.} It is important that the period of the FISDW potential is
not fixed rigidly, but varies with the magnetic field, so that the
electrons are redistributed between the Fermi ``reservoir'' below the
``primary'' FISDW gap and the ``Hall states'' above that gap. It is
because of this adjustment of the FISDW periodicity the Hall
conductivity maintains a constant value with the varying magnetic
field. This is in contrast to the models of Ref.\ \cite{Thouless},
where the periodicity of the potentials is fixed, and the Hall
conductivity jumps wildly when the magnetic field varies a little. In
the FISDW state, impurities are not necessary to produce the QHE,
except to pin the FISDW, because, as shown in the next Section, if the
FISDW is not pinned and is free to move, the QHE disappears.  The
(TMTSF)$_2$X materials seem to be the only substances where the QHE in
a 2D periodic potential is realized experimentally.

\section{Motion of the FISDW}
\label{Sec:Moving}

   In Sec.\ \ref{Sec:QHE} we have demonstrated that, when the FISDW is
pinned, the Hall conductivity is quantized. This result applies to the
case where the applied electric field is weak and time-independent.
On the other hand, when the electric field is strong or
time-dependent, the FISDW may move.  It is interesting to study how
this motion would influence the QHE. At first sight, since the
density-wave can move only along the chains, this purely 1D motion
cannot contribute to the Hall effect, which is essentially a 2D
effect. On the other hand, according to Eq.\ (\ref{jx}), the
Fr\"{o}hlich conductivity due to the motion of the FISDW does
contribute to the Hall current along the chains $j_x$ and, in this
way, may modify the Hall effect. To solve the problem, we need to find
how the velocity of the FISDW, $v_{\rm DW}\propto\dot{\Theta}$,
depends on $E_y$. It is well known that the electric field along the
chains, $E_x$, may induce the motion of a density wave along the
chains. However, it is not obvious whether the electric field {\em
across} the chains $E_y$ may induce the FISDW motion {\em along} the
chains. To study this issue, first we derive the equation of motion of
an ideal FISDW and then phenomenologically add pinning and damping of
the FISDW to the equation to make it more realistic. We stay within
the linear response theory, having in mind depinning of the FISDW by
an infinitesimal ac electric field, not by a strong dc field.  We
study rigid motion of the FISDW, where the phase $\Theta$ depends only
on the time $t$, but not on the coordinates $x$ and $y$. We restrict
consideration to the frequencies much lower than the FISDW gap and
take into account only collective motion of the FISDW, not
single-electron excitations across the gap.

   To derive the equation of motion of $\Theta(t)$, we need to know
the Lagrangian of the system, $L$.  Two terms in $L$ can be readily
recovered by taking into account that the current density $j_x$
(\ref{jx}) is the variational derivative of the Lagrangian with
respect to the electromagnetic vector-potential $A_x$:
\begin{equation}
j_x=c\,\delta L/\delta A_x.
\label{dL/dAx}
\end{equation}
Written in the gauge-invariant form, the recovered part of the
Lagrangian is equal to
\begin{equation}
L_1=-\sum_{i,j,k}\frac{Ne^2}{hc}\varepsilon_{ijk}A_iF_{jk}
-\frac{e}{\pi b}\Theta E_x,
\label{L1}
\end{equation}
where $\varepsilon_{ijk}$ is the antisymmetric tensor with the indices
$i,j,k=t,x,y$; $A_i$ and $F_{jk}$ are the vector-potential and the
tensor of the electromagnetic field; and $E_x\equiv F_{tx}$ is the
electric field along the chains.  The first term in Eq.\ (\ref{L1}) is
the so-called Chern--Simons term responsible for the QHE
\cite{Yakovenko91a}.  The second term describes the interaction of the
density-wave condensate with the electric field along the chains
\cite{Gruner94b}.

   Lagrangian (\ref{L1}) should be supplemented with the kinetic
energy of the FISDW condensate, $K$. Being produced by the
instantaneous Coulomb interaction between the electrons, the FISDW
potential itself has no inertia. So, $K$ consists of only the kinetic
energy of the electrons confined under the FISDW gap.  This energy is
proportional to the square of the average electron velocity, which, in
turn, is proportional to the electric current along the chains:
\begin{equation}
K=\frac{\pi\hbar b}{4v_{\rm F}e^2}\,j_x^2,                  \label{K}
\end{equation}
where $v_{\rm F}=k_{\rm F}/m$ is the Fermi velocity. Substituting Eq.\
(\ref{jx}) into Eq.\ (\ref{K}), expanding, and omitting the
unimportant term proportional to $E_y^2$, we obtain the second part of
the Lagrangian:
\begin{equation}
L_2=\frac{\hbar}{4\pi bv_{\rm F}}\dot{\Theta}^2
-\frac{eN}{2\pi v_{\rm F}}\dot{\Theta}E_y.                \label{L2}
\end{equation}
The first term in Eq.\ (\ref{L2}) is the same as the kinetic energy of
a purely 1D density wave \cite{Gruner94b} and is not specific to the
FISDW.  The most important is the second term which describes the
interaction of the FISDW motion and the electric field perpendicular
to the chains. This term is allowed by symmetry in the considered
system and has the structure of a mixed vector--scalar product:
\begin{equation}
{\bf v}_{\rm DW}[{\bf E}\times{\bf H}].
\label{vEH}
\end{equation}
Here, ${\bf v}_{\rm DW}$ is the velocity of the FISDW, which is
proportional to $\dot{\Theta}$ and is directed along the chains, that
is, along the $x$-axis.  The magnetic field ${\bf H}$ is directed
along the $z$-axis; thus, the electric field ${\bf E}$ may enter Eq.\
(\ref{vEH}) only through the component $E_y$.  Comparing formula
(\ref{vEH}) with the last term in Eq.\ (\ref{L2}), one should take
into account that the magnetic field enters the last term implicitly,
through the integer $N$, which depends on $H$ and changes sign when
$H$ changes sign.

   Varying the total Lagrangian $L=L_1+L_2$, given by Eqs.\ (\ref{L1})
and (\ref{L2}), with respect to $\Theta$, we find the equation of
motion of $\Theta(t)$:
\begin{equation}
\ddot{\Theta}=-\frac{2ev_{\rm F}}{\hbar}E_x + \frac{eNb}{\hbar}\dot{E_y}.
\label{EOM}
\end{equation}
In Eq.\ (\ref{EOM}), the first two terms constitute the standard 1D
equation of motion of the density wave \cite{Gruner94b}, whereas the
last term, proportional to the time derivative of $E_y$, comes from
the last term in Eq.\ (\ref{L2}) and describes the influence of the
electric field across the chains on the motion of the FISDW along the
chains.

   Setting $E_x=0$ and integrating Eq.\ (\ref{EOM}) in time, we find
that
\begin{equation}
\dot{\Theta}=eNbE_y/\hbar.
\label{theta.}
\end{equation}
Substituting Eq.\ (\ref{theta.}) into Eq.\ (\ref{jx}), we see that the
first term in Eq.\ (\ref{jx}) (the Fr\"{o}hlich conductivity of the
FISDW) precisely cancels the second term (the quantum Hall current),
so the total Hall current is equal to zero.  This result could have
been derived without calculations from the fact that the time
dependence $\Theta(t)$ is determined by the principle of minimal
action.  The relevant part of the action is given, in this case, by
Eq.\ (\ref{K}), which attains the minimal value at $j_x=0$. We can say
that, if the FISDW is free to move, it adjusts its velocity to
``compensate'' the external electric field $E_y$ and to keep zero Hall
current.

   It is instructive to see how the nullification of the Hall
conductivity takes place in the case where the electric field is
directed along the chains.  Varying $L$ (Eqs.\ (\ref{L1}) and
(\ref{L2})) with respect to $A_y$, we find the density of current
perpendicular to the chains:
\begin{equation}
j_y=-\frac{2Ne^2}{h}E_x-\frac{eN}{2\pi v_{\rm F}}\ddot{\Theta}.
\label{jy}
\end{equation}
In the r.h.s.\ of Eq.\ (\ref{jy}), the first term describes the
quantum Hall current, whereas the second term, proportional to the
{\it acceleration} of the FISDW condensate, comes from the last term
in Eq.\ (\ref{L2}) and reflects the contribution of the FISDW motion
along the chains to the electric current across the chains.  According
to the equation of motion (\ref{EOM}), the electric field along the
chains accelerates the density wave:
\begin{equation}
\ddot{\Theta}=-2ev_{\rm F}E_x/\hbar,
\label{theta..}
\end{equation}
thus, the Hall current (\ref{jy}) vanishes.

   However, it is clear that, in stationary, dc measurements, the
acceleration of the FISDW, discussed in the previous paragraph, cannot
last forever.  Any friction or dissipation will inevitably stabilize
the motion of the density wave to a steady flow with zero
acceleration.  In the steady state, the last term in Eq.\ (\ref{jy})
vanishes, and the current $j_y$ recovers its quantum Hall value.  The
same is true in the case where the electric field is perpendicular to
the chains.  In that case, the dissipation eventually stops the motion
of the FISDW along the chains and restores $j_x$ (\ref{jx}) to the
quantum Hall value. The conclusion is that the contribution of the
moving FISDW condensate to the Hall conductivity is essentially
nonstationary and cannot be observed in dc measurements.

\begin{figure}
\centerline{\psfig{file=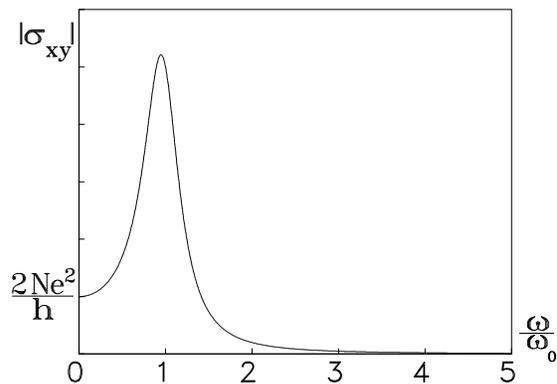,height=0.3\textheight,angle=-90}}
\caption{ Absolute value of the Hall conductivity in the FISDW state
as a function of the frequency $\omega$ normalized to the pinning
frequency $\omega_0$, as given by Eq.\ (\ref{omega}) with
$\omega_0\tau=2$.}
\label{Fig:sxy(w)}
\end{figure}

   On the other hand, the effect can be seen in ac measurements.  To
be realistic, let us add damping and pinning \cite{Gruner94b} to the
equation of motion of the FISDW (\ref{EOM}):
\begin{equation}
\ddot{\Theta}+\frac{1}{\tau}\dot{\Theta}+\omega_0^2\Theta
=-\frac{2ev_{\rm F}}{\hbar}E_x + \frac{eNb}{\hbar}\dot{E_y},
\label{fric}
\end{equation}
where $\tau$ is the relaxation time and $\omega_0$ is the pinning
frequency.  Solving Eq.\ (\ref{fric}) via the Fourier transformation
from the time $t$ to the frequency $\omega$ and substituting the
result into Eqs.\ (\ref{jx}) and (\ref{jy}), we find the Hall
conductivity as a function of the frequency:
\begin{equation}
\sigma_{xy}(\omega)=\frac{2Ne^2}{h}\frac{\omega_0^2-i\omega/\tau}
{\omega_0^2-\omega^2-i\omega/\tau}.
\label{omega}
\end{equation}
The absolute value of the Hall conductivity, $|\sigma_{xy}|$, computed
from Eq.\ (\ref{omega}) is plotted in Fig.\ \ref{Fig:sxy(w)} as a
function of $\omega/\omega_0$ for $\omega_0\tau=2$. The Hall
conductivity is quantized at zero frequency and has a resonance at the
pinning frequency.  At higher frequencies, where the pinning and the
damping can be neglected, and the system effectively behaves as an
ideal, purely inertial system considered above, the Hall conductivity
does decrease toward zero.

   Frequency dependence of the Hall conductivity in conventional,
semiconductor QHE systems was measured using the technique of crossed
wave guides \cite{Kuchar}, but no measurements have been done in a
FISDW system thus far. Such measurements would be very interesting,
because the ac behavior of the FISDW should differentiate the QHE in
(TMTSF)$_2$X from the conventional QHE in semiconductors. To give a
crude estimate of the required frequency range, we quote the value of
the pinning frequency $\omega_0\sim$ 3 GHz $\sim$ 0.1 K $\sim$ 10 cm
for a regular SDW (not FISDW) in (TMTSF)$_2$PF$_6$ \cite{Quinlivan}.

   Theoretically, frequency dependence of the Hall conductivity in a
FISDW system was considered in Ref.\ \cite{Maki89}.  This theory fails
to produce the QHE at zero frequency; thus, it does not agree with our
results.  The interplay between the QHE and the motion of the FISDW
was discussed in Ref.\ \cite{Rozhavsky92}.  Unfortunately, this paper
has troubles with calculations and physical interpretation and cannot
be considered as a consistent theory. The influence of the FISDW
motion on the QHE was described by the present authors in Ref.\
\cite{Yakovenko93c}.

   Due to the presence of the magnetic field in the problem, we could
phenomenologically add a term proportional to $E_y$ to Eq.\
(\ref{fric}) and a term proportional to $\dot{\Theta}$ to Eq.\
(\ref{jy}). These terms violate the time reversal symmetry of the
equations, which indicate the dissipative nature of these terms. Thus,
these terms cannot be derived within the Lagrangian formalism,
employed in this Section, and should be obtained from the Boltzmann
equation, where the time-reversal symmetry is already broken. Because
the dissipation is associated with the normal carriers thermally
excited across the FISDW energy gap, these terms should be
exponentially small and negligible at low temperatures. If taken into
account, these terms would modify the frequency dependence of
$\sigma_{xy}$ (Eq.\ (\ref{omega}) and Fig.\ \ref{Fig:sxy(w)}) at
intermediate frequencies, but not at zero and high frequencies.

   In this Section, we did not touch the issue of the FISDW depinning
by a strong dc electric field. In that case, the motion of the density
wave is controlled by dissipation, which is very difficult to study
theoretically on microscopic level. The influence of a steady motion
of a regular CDW on the Hall conductivity was studied theoretically in
Ref.\ \cite{Artemenko84b}.  The results can be interpreted in the
following way: The steady motion of the CDW condensate itself does not
contribute to the Hall effect; however, this motion influences the
thermally excited normal carriers and, in this way, affects the Hall
voltage.  This theory is complimentary to our theory, which studies
only the condensate contribution at zero temperature.  Mathematically,
the steady motion of the density wave modifies the Hall effect via the
dissipative terms discussed in the previous paragraph. Since the bare
value of the Hall conductivity in a regular CDW/SDW system is
determined by the normal carriers only, the steady motion of the
density wave produces a considerable, of the order of unity, effect on
the Hall conductivity, which was observed experimentally
\cite{Artemenko84a}.  On the other hand, in the case of the FISDW,
where the big quantum contribution from the electrons below the gap
dominates the Hall conductivity, the contribution of the thermally
excited normal carriers to the Hall conductivity should be negligible
at low temperatures. Thus, the steady motion of the FISDW should not
change considerably the Hall voltage, as, indeed, it was observed
experimentally in (TMTSF)$_2$ClO$_4$ \cite{Osada87}.  More recent
measurements in (TMTSF)$_2$PF$_6$ \cite{Balicas93} show results in
some sense opposite to the results of Ref.\ \cite{Osada87}.  The
origin of the difference is not clear at the moment.

\section{Finite Temperature}
\label{Sec:Temperature}

   The Hall conductivity at a finite temperature is not quantized
because of the presence of thermally excited quasiparticles above the
energy gap. It is interesting to find how the Hall conductivity
evolves with the temperature. Because the QHE at zero temperature is
generated by the collective motion of the electrons in the FISDW
condensate, the issue here is the temperature dependence of the
condensate current. Obviously, the condensate current must gradually
decrease with increasing temperature and vanish at the transition
temperature $T_c$, where the FISDW order parameter disappears.  This
behavior is qualitatively similar to the temperature dependence of the
superconducting condensate density and the inverse magnetic field
penetration depth in superconductors.

   In order to obtain explicit results, we need to make some
approximations. Let us linearize the parabolic dispersion law in
Hamiltonian (\ref{Hm}) near the Fermi energy:
\begin{equation}
\hbar^2 k_x^2/2m-E_{\rm F}\approx
\pm v_{\rm F}(k_x\mp k_{\rm F}),
\label{H-EF}
\end{equation}
and focus on the electrons whose momenta are close $+k_{\rm F}$ and
$-k_{\rm F}$. Let us count their momenta from $+k_{\rm F}$ and
$-k_{\rm F}$ and label their wave functions by the index $\pm$:
$\psi_+$ and $\psi_-$. In this representation, a complete electron
wave function is a spinor $(\psi_+,\psi_-)$, and the Hamiltonian is a
$2\times2$ matrix, which can be expanded over the Pauli matrices
$\hat{\tau}_1$, $\hat{\tau}_2$, $\hat{\tau}_3$, and the unity matrix
$\hat{1}$ (which we will not write explicitly in the following
formulas). Taking into account Eq.\ (\ref{Qx}), we can rewrite
Hamiltonian (\ref{Ham}) in the spinor representation as
\begin{equation}
\hat{\cal H}=-i\hbar v_{\rm F}\hat{\tau}_3\frac{\partial}{\partial x} +
\Delta \hat{\tau}_1 e^{i\hat{\tau}_3 (NG_xx-\Theta)}
+ 2t_b\cos[k_yb-G_x(x-v_{E_y}t)].
\label{Hlin}
\end{equation}
The last term in Eq.\ (\ref{Hlin}) can be eliminated by chiral
transformation of the electron wave function:\footnote{This kind of
transformation was first introduced in Ref.\ \cite{Lebed84} that
started development of the FISDW theory.}
\begin{equation}
\left( \begin{array}{c}\psi_+\\\psi_- \end{array} \right)
\:\rightarrow\;
\exp\left\{i\hat{\tau}_3 \frac{2t_b}{\hbar\omega_c}
\sin[k_yb-G_x(x-v_{E_y}t)]\right\}
\left( \begin{array}{c}\psi_+\\\psi_- \end{array} \right),
\label{->}
\end{equation}
where 
\begin{equation}
\hbar\omega_c=\hbar v_{\rm F}G_x=ebHv_{\rm F}/c
\label{wc}
\end{equation}
is the characteristic energy of the magnetic field (the cyclotron
frequency), which is equal to the distance in energy between the
Landau gaps discussed in Sec.\ \ref{Sec:FISDW}.  In representation
(\ref{->}), Hamiltonian (\ref{Hlin}) becomes
\begin{equation}
\hat{\cal H}=-i\hbar v_{\rm F}\hat{\tau}_3\frac{\partial}{\partial x} +
\Delta \hat{\tau}_1 \exp\{i\hat{\tau}_3 (NG_xx-\Theta)\}
\exp\left\{i\hat{\tau}_3 \frac{4t_b}{\hbar\omega_c}
\sin[k_yb-G_x(x-v_{E_y}t)]\right\}.
\label{Hnot}
\end{equation}
The chiral transformation (\ref{->}) has eliminated the hopping
potential from Hamiltonian (\ref{Hlin}) and transformed it into the
periodic function multiplying the FISDW potential in Eq.\
(\ref{Hnot}). Expanding that periodic function into the Fourier
series, we get the following expression:
\begin{equation}
\hat{\cal H}=-i\hbar v_{\rm F}\hat{\tau}_3\frac{\partial}{\partial x}
+ \Delta \hat{\tau}_1 e^{i\hat{\tau}_3 [N(k_yb+G_xv_{E_y}t)-\Theta]}
\sum_n a_{n+N}e^{i\hat{\tau}_3 n[k_yb-G_x(x-v_{E_y}t)]},
\label{Hsum}
\end{equation}
where the coefficients of the expansion, $a_n$, are the Bessel
functions:\footnote{General expression (\ref{Hsum}) is valid even when
the FISDW has a nonzero transverse wave vector and the transverse
dispersion law of the electrons is more complicated, but expression
(\ref{Bessel}) for the expansion coefficients $a_n$ would be different
in that case.}
\begin{equation}
a_n=J_n(4t_b/\hbar\omega_c).
\label{Bessel}
\end{equation}
The last term in Eq.\ (\ref{Hsum}) is the sum of many sinusoidal
potentials whose wave vectors are the integer multiples of the
magnetic wave vector $G_x$. Each of these periodic potentials mixes
the $+$ and $-$ electrons and opens an energy gap at the electron wave
vector $k_x$ shifted from $\pm k_{\rm F}$ by an integer multiple of
$G_x/2$. These multiple gaps are exactly the same gaps that were
discussed in Sec.\ \ref{Sec:FISDW}.

   The term with $n=0$ in the sum in Eq.\ (\ref{Hsum}) does not depend
on $x$ and opens a gap right at the Fermi level.\footnote{Since, by
introducing the $\pm$ electrons, we have already subtracted the wave
vectors $\pm k_{\rm F}$, the actual wave vector that corresponds to
this term is $2k_{\rm F}$.} When the temperature $T$ is much lower
than the distance between the energy gaps $\hbar\omega_c$:
\begin{equation}
T\ll\hbar\omega_c,
\label{T<<wc}
\end{equation}
only the gap at the Fermi level is important, whereas the other gaps
may be neglected.  Condition (\ref{T<<wc}) is always satisfied in the
relevant temperature range $0\leq T\leq T_c$ in the weak coupling
theory of the FISDW, where $T_c\ll\hbar\omega_c$.  Thus, let us omit
all the terms in the sum in Eq.\ (\ref{Hsum}), except the term with
$n=0$:
\begin{equation}
\hat{\cal H}=-i\hbar v_{\rm F}\hat{\tau}_3\frac{\partial}{\partial x}
+ \Delta_{\rm eff}\hat{\tau}_1 e^{i\hat{\tau}_3
[N(k_yb+G_xv_{E_y}t)-\Theta]},
\label{HaN}
\end{equation}
where
\begin{equation}
\Delta_{\rm eff}=a_N\Delta.
\label{Deff}
\end{equation}
This is the so-called single-gap approximation \cite{Montambaux86}.
As explained in Sec.\ \ref{Sec:FISDW}, in order to maximize the energy
gap at the Fermi level, the system selects such a value of $N$ that
maximizes the coefficient $a_N$ in Eq.\ (\ref{Deff}). It follows from
Eq.\ (\ref{Bessel}) and properties of the Bessel functions that the
maximum of $a_N$ is achieved at $N\approx 4t_b/\hbar\omega_c$
\cite{Lebed85}.\footnote{When the transverse wave vector of the FISDW
is not zero, the value of $N$ is controlled also by $t_b'$, the
next-nearest-chain hopping integral of electrons \cite{Montambaux85}.}
It was shown explicitly in Ref.\ \cite{Yakovenko91a} that omission of
the gaps located deeply below the Fermi energy does not change the
value of the Hall conductivity, at least at zero temperature.

   By the above sequence of manipulations, we have combined the two
periodic potentials in Eq.\ (\ref{Ham}) into the single effective
potential (\ref{HaN}) that opens a gap at the Fermi level. It follows
from Eq.\ (\ref{HaN}) that the phase $\varphi$ of this effective
potential changes in time at the rate proportional to the transverse
electric field $E_y$:
\begin{equation}
\dot{\varphi}=-NG_xv_{E_y},
\label{phi.}
\end{equation}
which means that the effective potential moves along the chains. Since
all electrons are below the energy gap opened by this potential, the
motion of the potential induces the Fr\"ohlich current along the
chains:
\begin{equation}
j_x=-\frac{e}{\pi b}\dot{\varphi}.
\label{jxphi}
\end{equation}
Substituting Eqs.\ (\ref{phi.}), (\ref{Gx}), and (\ref{vEy}) into Eq.\
(\ref{jxphi}), we find the QHE in agreement with Eq.\ (\ref{2Ne2/h}):
\begin{equation}
j_x=\frac{2Ne^2}{h}E_y.
\label{jxEy}
\end{equation}
To avoid confusion, we wish to emphasize that, unlike in Sec.\
\ref{Sec:Moving}, here the FISDW is assumed to be immobile, and the
FISDW phase $\Theta$ in Eq.\ (\ref{HaN}) is time-independent. The
effective potential (\ref{HaN}) moves, because it is a combination of
the stationary FISDW potential and the moving hopping potential
(\ref{Ham}).

   Eq.\ (\ref{jxphi}) is a good starting point to discuss the
temperature dependence of the QHE. According to the above
consideration, the Hall conductivity is the Fr\"ohlich conductivity of
the effective periodic potential (\ref{HaN}). Thus, the temperature
dependence of the QHE must be the same as the temperature dependence
of the Fr\"ohlich conductivity. The latter issue was studied in the
theory of the CDW \cite{Lee79,Maki90}. At a finite temperature $T$,
the electric current carried by the CDW condensate is reduced with
respect to the zero-temperature value by a factor $f(T)$. The same
factor reduces the condensate Hall effect at a finite temperature:
\begin{equation}
\sigma_{xy}(T)=f(T)\,2Ne^2/h,
\label{f(T)2Ne2/h}
\end{equation}
\begin{equation}
f(T)=1-\int_{-\infty}^\infty \frac{dk_x}{\hbar v_{\rm F}}
\left(\frac{\partial E}{\partial k_x}\right)^2
\left[-\frac{\partial n_{\rm F}(E/k_{\rm B}T)}{\partial E}\right],
\label{f(T)}
\end{equation}
where $E=\sqrt{(\hbar v_{\rm F}k_x)^2+\Delta_{\rm eff}^2}$ is the
electron dispersion law in the FISDW phase, $k_{\rm B}$ is the
Boltzmann constant, and $n_{\rm F}(\epsilon)=(e^\epsilon+1)^{-1}$ is
the Fermi distribution function.  The last term in Eq.\ (\ref{f(T)})
reflects the fact that normal quasiparticles, thermally excited above
the energy gap, equilibrate with the immobile crystal lattice; thus,
only a fraction of all electrons is carried along the chains by the
moving periodic potential, which reduces the Hall/Fr\"ohlich current.
A simple, transparent derivation of Eq.\ (\ref{f(T)}) is given in
Ref.\ \cite{Yakovenko96b}.

\begin{figure}
\centerline{\psfig{file=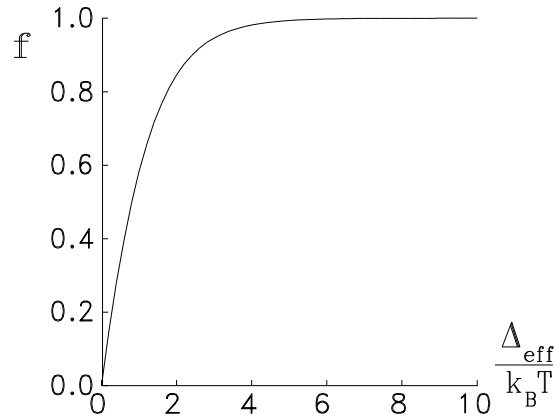,height=0.3\textheight,angle=-90}}
\caption{ The reduction factor $f$ of the Hall conductivity as a
function of the ratio of the energy gap at the Fermi level
$\Delta_{\rm eff}$ to the temperature $T$, as given by Eq.\
(\ref{dynMaki}).}
\label{Fig:f(T)}
\end{figure}

\begin{figure}
\centerline{\psfig{file=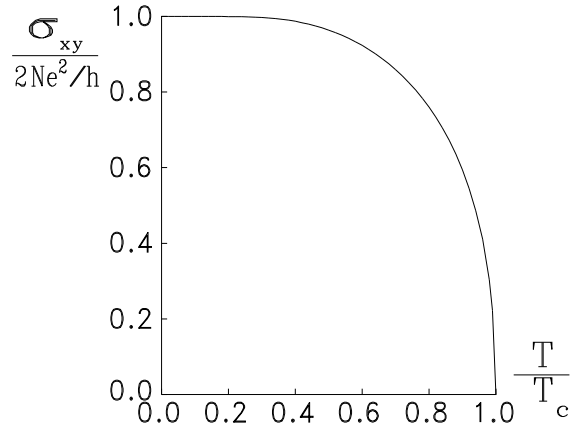,height=0.3\textheight,angle=-90}}
\caption{ Hall conductivity in the FISDW state as a function of the
temperature $T$ normalized to the FISDW transition temperature $T_c$.}
\label{Fig:sxy(T)}
\end{figure}

   The function $f$ (\ref{f(T)}) depends only on the ratio of the
energy gap at the Fermi level, $\Delta_{\rm eff}$ (\ref{Deff}), and
the temperature $T$ and can be written as \cite{Maki89,Maki90}
\begin{equation}
f\left(\frac{\Delta_{\rm eff}}{k_{\rm B}T}\right)=
\int_0^\infty d\zeta\,
\tanh\left(\frac{\Delta_{\rm eff}}{2k_{\rm B}T}\cosh\zeta\right)
/\cosh^2\zeta.
\label{dynMaki}
\end{equation}
The function $f$ is plotted in Fig.\ \ref{Fig:f(T)}. It is equal to 1
at zero temperature, where Eq.\ (\ref{f(T)2Ne2/h}) gives the QHE,
gradually decreases with increasing $T$, and vanishes when
$T\gg\Delta_{\rm eff}$. Taking into account that the FISDW order
parameter $\Delta$ itself depends on $T$ and vanishes at the FISDW
transition temperature $T_c$, it is clear that $f(T)$ and
$\sigma_{xy}(T)$ vanish at $T\rightarrow T_c$, where
$\sigma_{xy}(T)\propto
f(T)\propto\Delta(T)\propto\sqrt{T_c-T}$. Assuming that the
temperature dependence $\Delta_{\rm eff}(T)$ is given by the BCS
theory \cite{Montambaux86}, we plot the temperature dependence of the
Hall conductivity, $\sigma_{xy}(T)$, in Fig.\ \ref{Fig:sxy(T)}.
Strictly speaking, Eq.\ (\ref{f(T)2Ne2/h}) gives only the Hall effect
of the FISDW condensate and should be supplemented with the Hall
conductivity of the thermally excited normal carriers. Then, at
$T\rightarrow T_c$, $\sigma_{xy}(T)$ should not vanish, but approach
to the Hall conductivity of the metallic phase. The latter is
determined by the distribution of the electron scattering time over
the Fermi surface and is small. The curve shown in Fig.\
\ref{Fig:sxy(T)} should be modified accordingly in a small vicinity of
$T_c$.

   The function $f(T)$ (\ref{f(T)}) is qualitatively similar to the
function $f_{\rm s}(T)$ that describes the temperature reduction of
the superconducting condensate density in the London case. Both
functions approach 1 at zero temperature, but near $T_c$ the
superconducting function behaves differently: $f_{\rm s}(T)\propto
\Delta^2(T)\propto T_c-T$. To understand the origin of the difference
between the two functions, one should consider them at a small, but
finite frequency $\omega$ and wave vector $q$.  Eqs.\ (\ref{f(T)}) and
(\ref{dynMaki}) represent the limit where $q/\omega=0$. This is the
relevant limit in our case, because the electric field is, supposedly,
strictly homogeneous in space ($q=0$), but may be time-dependent
($\omega\neq0$). The effective periodic potential (\ref{HaN}) is also
time-dependent. On the other hand, for the Meissner effect in
superconductors, where the magnetic field is stationary ($\omega=0$),
but varies in space ($q\neq0$), the opposite limit $\omega/q=0$ is
relevant. That is why $f(T)$ and $f_{\rm s}(T)$ are different.

   The function $f(T)$ for the Fr\"ohlich current of a regular CDW/SDW
was calculated in Ref.\ \cite{Lee79} in the form (\ref{f(T)}) and in
Refs.\ \cite{Maki89,Maki90} in the form (\ref{dynMaki}).  The Hall
conductivity in the FISDW state at a finite temperature was discussed
in Ref.\ \cite{Maki89}, which failed to produce the QHE at zero
temperature. Temperature dependence of the Hall resistance in
(TMTSF)$_2$X was measured in experiments \cite{Chaikin92e}. However,
to compare the experimental results with our theory, it is necessary
to convert the Hall resistivity into the Hall conductivity, which
requires experimental knowledge of all components of the resistivity
tensor.

   Because the FISDW phase $\Theta$ enters linearly into the phase
$\varphi$ of the effective periodic potential in Eq.\ (\ref{HaN}), the
results of Sec.\ \ref{Sec:Moving} could be immediately generalized to
a finite temperature. When the FISDW moves and its phase $\Theta$
depends on time, the r.h.s.\ of Eq.\ (\ref{jx}) should be multiplied
by the function $f(T)$. The frequency-dependent Hall conductivity,
given by Eq.\ (\ref{omega}) and shown in Fig.\ \ref{Fig:sxy(w)},
should be also multiplied by $f(T)$.\footnote{The phenomenological
parameters $\tau$ and $\omega_0$ may also depend on temperature.}
Such a simple generalization of the results of this Section to finite
frequencies would be possible, because the function $f$ has no
frequency dependence for $\omega\ll\Delta_{\rm eff}$. However, at a
finite temperature, the dissipative terms, discussed at the end of
Sec.\ \ref{Sec:Moving}, may become comparable with the other terms and
significantly change $\sigma_{xy}(\omega)$ beyond multiplication by
the factor $f(T)$.

\section{Edge States}
\label{Sec:Edges}

   Thus far we treated the QHE as a bulk phenomenon and did not pay
attention to the edges of the crystal. On the other hand, it is known
that the theory of the QHE can be reformulated in terms of the gapless
edge states located at the boundaries of a Hall sample
\cite{Halperin82}. The edge states in (TMTSF)$_2$X attracted attention
in recent studies of the chiral states on the surface of a bulk QHE
sample \cite{Chalker95}.  Let us show how the QHE in the FISDW state
can be formulated in terms of the edge states. We will consider a
sample that is infinite in the $x$ direction along the chains and has
a finite macroscopic size $2L_y$ in the $y$ direction across the
chains: $-L_y\leq y\leq L_y$. The edge states are located near the
boundaries of the sample at $y=\pm L_y$. The total number of the
chains in the crystal, $M_{\rm max}$, is finite: $M_{\rm max}=2L_y/b$.

   To introduce the edge states in a most natural way, let us
reformulate the FISDW picture using the Wannier representation of the
electron wave functions \cite{Yakovenko87b}. First, let us find the
electron eigenfunctions in the metallic state, in the absence of the
FISDW. The Schr\"odinger equation that corresponds to Hamiltonian
(\ref{Hlin}) with $\Delta=0$ and $E_y=0$,
\begin{equation}
[\mp i\hbar v_{\rm F}\frac{\partial}{\partial x}+2t_b\cos(k_yb-G_xx)]
\psi_{k_x,k_y,\pm}(x)=\varepsilon\psi_{k_x,k_y,\pm}(x),
\label{Schrodinger}
\end{equation}
has the following solution:
\begin{eqnarray}
\psi_{k_x,k_y,\pm}(x,n,t)&=&\exp\{i[-\frac{\varepsilon t}{\hbar}+
k_xx+k_ynb \pm\frac{2t_b}{\hbar\omega_c}\sin(k_yb-G_xx)]\},
\label{psi+-} \\
\varepsilon&=&\pm\hbar v_{\rm F}k_x.
\label{epsilon}
\end{eqnarray}
In Eq.\ (\ref{psi+-}), the wave vectors $k_x$ and $k_y$ are the
quantum numbers that label the energy eigenfunctions, whereas $x$ and
$n=y/b$ are the running coordinates of the wave functions. Note that
the dispersion law (\ref{epsilon}) is purely 1D: The energy
$\varepsilon$ depends on $k_x$, but does not depend on $k_y$. As
mentioned in Sec.\ \ref{Sec:FISDW}, this is a consequence of the
Landau degeneracy in magnetic field. Because of the degeneracy in
$k_y$, any superposition of eigenstates (\ref{psi+-}) with different
$k_y$ also is an energy eigenstate. Let us superimpose functions
(\ref{psi+-}) with the coefficients of the Fourier transform:
\begin{eqnarray}
\psi_{k_x,M,\pm}(x,n,t)&=&
b\int\frac{dk_y}{2\pi}\,\psi_{k_x,k_y,\pm}(x,n,t)\,e^{-ik_yMb}
\nonumber\\
&=& e^{i[-\varepsilon t/\hbar+k_xx+(n-M)G_xx]}
J_{n-M}(\mp 2t_b/\hbar\omega_c),
\label{JnM}
\end{eqnarray}
where $J_n(\xi)$ is the Bessel function of the $n$-th order.

   The Wannier wave functions (\ref{JnM}) form a new complete set of
the energy eigenfunctions. These functions are delocalized along the
chains, because they are the plane waves in the $x$ direction. The
shape of the wave functions across the chains is given by the Bessel
function $J_n(2t_b/\hbar\omega_c)$ considered as a function of its
index $n$ with the fixed argument $2t_b/\hbar\omega_c$, which is the
ratio of the hopping integral between the chains to the cyclotron
frequency of the magnetic field.  The Bessel function
$J_n(2t_b/\hbar\omega_c)$ has a maximum at $n\approx
2t_b/\hbar\omega_c$ and exponentially decreases to zero as $n$
increases further.  Thus, the wave functions (\ref{JnM}) are localized
across the chains with the characteristic width $4t_b/\hbar\omega_c$,
which decreases with increasing magnetic field $H$ as $1/H$.  Each
wave function (\ref{JnM}) is centered on a certain chain labeled by
the quantum number $M$.

   The wave functions (\ref{JnM}) are qualitatively similar to the
Landau wave functions of an isotropic particle in magnetic field. The
both sets of the wave functions are localized in one direction and
delocalized in another, and the energy does not depend on the position
where the localized wave function is placed. However, because our
problem is strongly anisotropic, the shapes of the wave functions are
different: the Bessel function in our case\footnote{If the transverse
dispersion law of the electrons is more complicated, the shape of the
wave function may differ from the Bessel function, but all qualitative
features of the Wannier functions, such as the localization across the
chains, remain valid.} and the Gaussian function in the Landau case.

   Since the wave functions $\psi_{k_x,M,\pm}$ form a complete basis,
we can use this Wannier basis to describe our system.  Let us
introduce the operators $\hat{a}^+_\pm(k_x,M)$ and
$\hat{a}_\pm(k_x,M)$ that create and annihilate an electron on a
Wannier chain $M$ in the state $\psi_{k_x,M,\pm}$.  Now, let us take
into account the FISDW potential $2\Delta\cos(Q_x x)$ in Eq.\
(\ref{HH}) with the wave vector (\ref{Qx}).  The matrix elements of
the FISDW potential between the states (\ref{JnM}) can be easily
evaluated. Keeping only the term that opens an energy gap at the Fermi
level, we get the following expression for Hamiltonian (\ref{HH}) in
the Wannier basis:
\begin{eqnarray}
  \hat{\cal H}&=&\int\frac{dk_x}{2\pi}\sum_M\,
  v_{\rm F}k_x[\hat{a}^+_+(k_x,M)\hat{a}_+(k_x,M)
  -\hat{a}^+_-(k_x,M)\hat{a}_-(k_x,M)]
  \nonumber \\
&&
  {}+\Delta_{\rm eff}[\hat{a}^+_+(k_x,M+N)\hat{a}_-(k_x,M)
  +\hat{a}^+_-(k_x,M)\hat{a}_+(k_x,M+N)],
\label{Hint}
\end{eqnarray}
where $\Delta_{\rm eff}=\Delta J_{N}(4t_b/\hbar\omega_c)$ is the same
as in Eqs.\ (\ref{Deff}) and (\ref{Bessel}).  There is no
single-electron hopping between the Wannier chains in Hamiltonian
(\ref{Hint}), but the FISDW potential scatters the $-$ electrons into
the $+$ electrons and simultaneously displaces them across the chains
by $N$ Wannier chains, where $N$ is the parameter of the FISDW.  In
the Wannier representation, it is very transparent why many different
FISDWs are possible in our 2D system in magnetic field.  In a purely
1D case, a CDW/SDW may couple the $+$ and $-$ electrons only on the
same chain.  In our 2D system, the FISDW may couple the $+$ and $-$
electrons on different chains; thus, the FISDW is characterized by the
integer distance $N$ between the coupled chains, which may take many
different values.

   The FISDW potential in Eq.\ (\ref{Hint}) hybridizes the $-$
electrons on the Wannier chain $M$ and the $+$ electrons on the
Wannier chain $M+N$ and opens a gap at the Fermi level in their energy
spectrum.  That procedure works for the chains in the bulk of the
crystal.  However, the states at the edges of the crystal are
exceptional.  The $+$ electron on the first $N$ chains on one side of
the crystal and the $-$ electrons on the last $N$ chains on the other
side of the crystal have no partner chains to couple with, so these
electrons remain ungapped.  Thus, the one side of the sample possesses
$N$ gapless chiral modes propagating along the edge with the velocity
$v_{\rm F}$, and the other side has $N$ gapless chiral modes
propagating in the opposite direction with the velocity $-v_{\rm F}$.

   Now, let us discuss the QHE in this system.  Suppose a small
electric voltage $V_y$ is applied across the chains.  That means that
the chemical potential varies across the chains.  Because all states
in the bulk of the crystal are gapped out, they would not respond to
this perturbation.  However, since the edge modes are not gapped, the
difference of the chemical potentials between the two edges produces
an imbalance between the occupation numbers of the modes at the
opposite edges, $\delta\rho\propto V_y$, which generates a net current
$I_x$ along the chains:
\begin{equation}
I_x=ev_{\rm F}N\delta\rho=ev_{\rm F}N\frac{2eV_y}{2\pi\hbar v_{\rm F}}
=\frac{2Ne^2}{h}V_y.
\label{Ix}
\end{equation}
Eq.\ (\ref{Ix}) represents the QHE, this time for the Hall
conductance, rather than conductivity (\ref{2Ne2/h}), which coincide
in 2D.

   The above derivation might have produced impression that the Hall
current flows only along the edges of the sample and is zero in the
bulk.  That is not necessarily the case.  Let us show how the bulk and
the edge pictures of the QHE connect with each other.  Suppose the
applied voltage $V_y$ drops homogeneously across the chains, so that
there is a tiny voltage drop $V_y/M_{\rm max}$ between every pair of
neighboring chains.  Because of the variation of the chemical
potential across the chains, the electron concentration and, thus, the
Fermi momentum $k_{\rm F}$ must change from chain to chain.  That
creates a problem when the FISDW pairs the $+$ and $-$ electrons on
different chains, where the Fermi momenta may be different.  When the
FISDW is pinned and does not move, which we assume to be the case
here, the states paired by the FISDW must have exactly opposite
momenta.  (If the paired momenta are different in absolute values, the
total momentum of the electrons under the gap is not zero, which means
that the FISDW moves.)  So, the momentum distribution of the electrons
on each chains must shift in $k_x$ to make $-k_{\rm F}$ at the chain
$M$ equal to $+k_{\rm F}$ at the chain $M+N$.  Since the momentum
distribution on each chain is shifted away from the symmetric
position, each chain carries an electric current.

   Let us illustrate this reasoning quantitatively.  The current on a
chain $M$ is the difference of the current $I^+_M$ carried by the
electrons with the positive momenta and the current $I^-_M$ of the
electrons with the negative momenta.  So, the total current is
\begin{equation}
I_x=
\begin{array}[t]{@{+\hspace{\arraycolsep}}l@{-\hspace{\arraycolsep}}l}
I^+_1 & I^-_1 \\
I^+_2 & I^-_2 \\
\multicolumn{2}{c}{\cdots} \\
I^+_{M_{\rm max}} & I^-_{M_{\rm max}}.
\end{array}
\label{I+-}
\end{equation}
Each line in this equation represents the current on a given chain.
As it was explained above, because the chains $M$ and $M+N$ are
coupled by the pinned FISDW, we have
\begin{equation}
I^-_M=I^+_{M+N}.
\label{IMN}
\end{equation}
Substituting Eq.\ (\ref{IMN}) into Eq.\ (\ref{I+-}), we find that the
total current is the difference of the edge currents:
\begin{equation}
I_x=\sum_{M=1}^N I^+_M-\sum_{M=M_{\rm max}-N+1}^{M_{\rm max}} I^-_M.
\label{IN}
\end{equation}
At the same time, the current on a given chain $M$ is not zero:
$I^+_M-I^-_M\neq0$.  That means that the total current (\ref{Ix}),
whose value is given by the difference of the edge terms (\ref{IN}),
is spread homogeneously over all chains, so that each chain carries a
portion of the total current.

   It is easy to see that the Hall current flows in those regions of
the crystal where the transverse voltage drops.  The total Hall
current is always given by Eq.\ (\ref{Ix}); nevertheless, the actual
physical distribution of the Hall voltage and current may be of some
interest.  In the semiconductor QHE devices, the experiment seems to
indicate that all Hall voltage drops near the sample boundaries, but
no such measurements were performed in the (TMTSF)$_2$X materials.

   If the FISDW is allowed to move, it is not required to pair the
exactly opposite electron momenta. Then, instead of (\ref{IMN}), we can
have $I^+_M=I^-_M$, so the current on each chain and the total current
(\ref{I+-}) are equal to zero.  This is in agreement with the result
of Sec.\ \ref{Sec:Moving} that, if the FISDW is free, there is no Hall
effect.

   Strictly speaking, expression (\ref{JnM}) for the Wannier functions
is valid only in the bulk and should be modified near the
edges. Although we neglected such complications in the above
discussion, we believe that our qualitative results would remain valid
in a more accurate theory.

\section{Conclusions}
\label{Sec:Conclusion}

   Main results reviewed in this paper can be summarized as
follows. When a 2D system that consists of parallel conducing chains
is placed in a strong magnetic field, the magnetic-field-induced
spin-density wave (FISDW) may appear in the system. The FISDW couples
the electron states at different chains, thus it is characterized by
an integer number $N$, the distance between the coupled chains. By
hybridizing the electron states at the opposite sides of the Fermi
surface, the FISDW opens an energy gap at the Fermi level everywhere
in the bulk of the crystal. However, on the $N$ chains at the both
edges of the crystal, half of the electron states remain ungapped,
because they have no partner chains to couple with. The electrons in
these $N$ chiral gapless modes propagate with the opposite velocities
at the opposite edges.

   When an electric field is applied, at zero temperature, the system
exhibits the quantum Hall effect (\ref{2Ne2/h}) with the same integer
number $N$ that characterizes the FISDW. As the temperature increases,
the Hall conductivity decreases, vanishing at the FISDW transition
temperature $T_c$. The function $f(T)$ that describes the reduction of
the Hall effect with the temperature is the same as the temperature
reduction function of the Fr\"ohlich current of a regular
charge/spin-density wave.  If the Hall effect is measured at a high
enough frequency, the motion of the FISDW produces an additional
contribution to the Hall current, such that the total Hall
conductivity vanishes.

   This work was partially supported by the NSF under Grant
DMR--9417451, by the Alfred P.~Sloan Foundation, and by the David and
Lucile Packard Foundation.


\begin{thebibliography}{99}
\itemsep 0pt

\bibitem{Yamaji90} Ishiguro T. and Yamaji K., Organic Superconductors
(Springer-Verlag, Berlin, 1990) Chapter 9.

\bibitem{Chaikin88d} Chamberlin R.V. {\it et~al.}, {\em Synth. Met.}
{\bf 27} (1988) B41.

   \bibitem{Pesty85} Pesty F., Garoche P., and Bechgaard K., {\em Phys.
Rev. Lett.} {\bf 55} (1985) 2495; Fortune N.A. {\it et~al.}, {\em ibid.}
{\bf 64} (1990) 2054.

\bibitem{Chaikin85} Naughton M.J. {\it et~al.}, {\em Phys.  Rev.
Lett.} {\bf 55} (1985) 969.

\bibitem{Takahashi84} Takahashi T., J\'erome D., and Bechgaard K.,
{\em J. Phys. France} {\bf 45} (1984) 945.

   \bibitem{Jerome89} Cooper J.R. {\it et~al.}, {\em Phys.  Rev. Lett.}
{\bf 63} (1988) 1984; Hannahs S.T. {\it et~al.}, {\em ibid.} {\bf 63}
(1988) 1988.

   \bibitem{Chaikin88a} Chamberlin R.V. {\it et~al.}, {\em Phys.  Rev.
Lett.} {\bf 60} (1988) 1189; Naughton M.J. {\it et~al.}, {\em ibid.}
{\bf 61} (1988) 621.

\bibitem{Jerome91b} Kang W., Cooper J.R., and J\'{e}rome D., {\em
Phys. Rev. B} {\bf 43} (1991) 11467.

\bibitem{Montambaux91} Montambaux G., {\em Physica Scripta T} {\bf 35}
(1991) 188.

\bibitem{Montambaux84b} H\'eritier M., Montambaux G., and Lederer P.,
{\em J. Phys. Lett. (Paris)} {\bf 45} (1984) L943.

\bibitem{Chaikin87a} Azbel M.Y., Bak P., and Chaikin P.M., {\em
Phys. Rev. Lett.} {\bf 59} (1987) 926.

\bibitem{Montambaux87} Poilblanc D. {\it et~al.}, {\it Phys.  Rev.
Lett.} {\bf 58} (1987) 270.

\bibitem{Yakovenko91a} Yakovenko V.M., {\it Phys. Rev. B} {\bf 43}
(1991) 11353.

\bibitem{Yakovenko94a} Machida K. {\it et~al.}, {\it Phys.  Rev. B}
{\bf 50} (1994) 921.

\bibitem{Lebed90} Lebed' A.G., {\em Pis'ma Zh. Exp. Teor. Fiz.} {\bf
51} (1990) 583 ({\em JETP Lett.} {\bf 51} (1990) 663); Machida K. and
Nakano M., {\em J. Phys. Soc. Jpn.} {\bf 59} (1990) 4223.

\bibitem{Gruner94b} Gr\"uner G., Density Waves in Solids
(Addison-Wesley, New York, 1994).

\bibitem{Montambaux86} Poilblanc D. {\it et~al.}, {\em J. Phys. C}
{\bf 19} (1986) L321; Virosztek A., Chen L., and Maki K., {\em
Phys. Rev. B} {\bf 34} (1986) 3371.

\bibitem{Lebed85} Lebed' A.G., {\em Zh. Exp. Teor. Fiz.} {\bf 89}
(1985) 1034 ({\em Sov. Phys. JETP} {\bf 62} (1985) 595).

\bibitem{Montambaux85} Montambaux G., H\'{e}ritier M., and Lederer P.,
{\em Phys. Rev. Lett.} {\bf 55} (1985) 2078; Montambaux G. and Zanchi
D., {\em ibid.} {\bf 77} (1996) 366.

\bibitem{Zak85} Dana I., Avron Y., and Zak J., {\em J. Phys. C} {\bf
18} (1985) L679.

\bibitem{Kunz86} Kunz H., {\em Phys. Rev. Lett.} {\bf 57} (1986) 1095;
Kohmoto M., {\em J. Phys. Soc. Jpn.} {\bf 62} (1993) 659.

\bibitem{Laughlin81} Laughlin R.B., {\it Phys. Rev. B} {\bf 23} (1981)
5632.

\bibitem{Thouless} Thouless D.J., in The Quantum Hall Effect,
R.E. Prange and S.M. Girvin, Eds. (Springer-Verlag, Berlin, 1987)
p. 101; Kohmoto M., {\em Phys. Rev. B} {\bf 39} (1989) 11943.

\bibitem{Kuchar} Kuchar F. {\it et~al.}, {\it Phys. Rev. B} {\bf 33}
(1986) 2965; Galchenkov L.A. {\it et~al.}, {\it Pis'ma v
Zh. Exp. Teor.  Fiz.} {\bf 46} (1987) 430 [{\it JETP Lett.}  {\bf 46}
(1987) 542].

\bibitem{Quinlivan} Quinlivan D. {\it et~al.}, {\it Phys.  Rev.
Lett.} {\bf 65} (1990) 1816.

\bibitem{Artemenko84b}
Artemenko S.~N. and Kruglov A.~N., {\it Fiz.  Tverd. Tela} {\bf 26}
   (1984) 2391 [{\it Sov. Phys. Solid State} {\bf 26} (1984) 1448];
Dolgov E.~N., {\it Fiz. Nizk. Temp.} {\bf 10} (1984) 911 [{\it
   Sov. J. Low Temp. Phys.} {\bf 10} (1984) 747].

\bibitem{Artemenko84a}
Artemenko S.~N. {\it et~al.}, {\it Pis'ma v Zh. Exp. Teor.  Fiz.} {\bf
   39} (1984) 258 [{\it JETP Lett.} {\bf 39} (1984) 308];
Forr\'{o} L. {\it et~al.}, {\it Phys.  Rev. B} {\bf 34} (1986) 9047;
Forr\'{o} L. {\it et~al.}, {\it Sol. State Comm.} {\bf 62} (1987) 715;
Tr\ae tteberg O., Balicas L., and Kriza G., {\it J. Phys. IV France,
   Colloque C2} {\bf 3} (1993) 61.

\bibitem{Osada87} Osada T. {\it et~al.}, {\it Phys. Rev. Lett.} {\bf
58} (1987) 1563.

\bibitem{Balicas93} Balicas L., Bi\v{s}kup N., and Kriza G., {\it
J. Phys. IV France, Colloque C2} {\bf 3} (1993) 319.

\bibitem{Maki89} Virosztek A. and Maki K., {\it Phys. Rev. B} {\bf 39}
(1989) 616.

\bibitem{Rozhavsky92} Petrova T. G. and Rozhavsky A. S., {\it Fiz.
Nizk.  Temp.} {\bf 18} (1992) 987 and {\it J. Phys. IV France,
Colloque C2} {\bf 3} (1993) 303.

\bibitem{Yakovenko93c} Yakovenko V.M., {\it J. Phys. IV France,
Colloque C2} {\bf 3} (1993) 307; {\it J. Supercond.} {\bf 7} (1994)
683; Yakovenko V.M. and Goan H.-S., in Proceedings of the Physical
Phenomena at High Magnetic Fields--II Conference, Z. Fisk {\it
et~al.}, Eds.  (World Scientific, Singapore, 1996) p. 116.

\bibitem{Lebed84} Gor'kov L.P. and Lebed' A.G., {\em
J. Phys. Lett. (Paris)} {\bf 45} (1984) L433.

\bibitem{Lee79} Lee P.A. and Rice T.M., {\it Phys. Rev. B} {\bf 19}
(1979) 3970; Rice T.M., Lee P.A., and Cross M.C., {\it ibid.} {\bf 20}
(1979) 1345.

\bibitem{Maki90} Maki K. and Virosztek A., {\it Phys. Rev. B} {\bf 41}
(1990) 557; {\bf 42} (1990) 655.

\bibitem{Yakovenko96b} Goan H.-S. and Yakovenko V.M.,
cond-mat/9607199, to be published in {\em Synth. Met.}

\bibitem{Chaikin92e} Kang W. {\it et~al.}, {\it Phys. Rev. B} {\bf 45}
(1992) 13566; see also Valfells S. {\it et~al.}, cond-mat/9606212.

\bibitem{Halperin82} Halperin B.I., {\it Phys. Rev. B} {\bf 25} (1982)
2185.

\bibitem{Chalker95} Chalker J.T. and Dohmen A., {\em Phys. Rev. Lett.}
{\bf 75} (1995) 4496; Balents L. and Fisher M.P.A., {\it ibid.}  {\bf
76} (1996) 2782.

\bibitem{Yakovenko87b} Yakovenko V.M., {\em Zh. Exp. Teor. Fiz.} {\bf
93} (1987) 627 ({\em Sov. Phys. JETP} {\bf 66} (1987) 355) and {\em
Europhys. Lett.} {\bf 3} (1987) 1041.

\end{thebibliography}
\end{document}